\documentclass[pre,aps,showkeys,showpacs,twocolumn,amsmath,amssymb]{revtex4}
\usepackage{dsfont,graphicx,psfrag}
\usepackage[mathscr]{euscript}
\psfrag{Pi}{$\pi$}
\psfrag{2Pi}{$2\pi$}
\psfrag{t}{$t$}
\psfrag{lambda}{$\lambda$}
\psfrag{D}{$D$}
\psfrag{Delta lambda}{$\Delta\lambda$}

\begin{document}
\title{Computing the optimal protocol for finite-time processes in
  stochastic thermodynamics} \author{Holger Then and Andreas Engel}
\affiliation{Institut f\"ur Physik, Universit\"at Oldenburg, 26111
  Oldenburg, Germany}

\begin{abstract}
  Asking for the optimal protocol of an external control parameter
  that minimizes the mean work required to drive a nano-scale system
  from one equilibrium state to another in finite time, Schmiedl and
  Seifert ({\it Phys. Rev. Lett.} {\bf 98}, 108301 (2007)) found the
  Euler-Lagrange equation to be a non-local integro-differential
  equation of correlation functions. For two linear examples, we show
  how this integro-differential equation can be solved analytically.
  For non-linear physical systems we show how the optimal protocol can
  be found numerically and demonstrate that there may exist several
  distinct optimal protocols simultaneously, and we present optimal
  protocols that have one, two, and three jumps, respectively.
\end{abstract}


\maketitle

\section{Introduction}
There exist plenty of reasons for processes to be optimized.  Both,
economically and ecologically, it is of high interest to minimize the
energy consumption.  Mathematically, the issue addresses the question
of designing an optimal protocol $\lambda(t)$ according to which some
dynamical system is driven from a given initial state to some other
desired final state.  Macroscopically, from the second law of
thermodynamics, there is a lower bound on the required work.  If
started from thermal equilibrium, the applied work needed to reach the
final state in an isothermal process is always larger than or equal to
the free energy difference.  The amount of the applied work that
exceeds the free energy difference is called the dissipated work and
is lost by heating the environment.  Since tasks need to be finished
within a given finite amount of time, the dissipated work is always
positive and depends on the details of the protocol resulting in a
technical challenge of how to prevent powerful engines and fast
microprocessors from their heat death.

Along with miniaturization new aspects arise. Microscopic systems are
subject to both, deterministic and stochastic forces, and it becomes
necessary to consider ensemble averages. To ensure that a quantum
computer is in a well defined state, the fluctuations have to be
minimized. Finding the optimal protocol that yields the least
fluctuations (beyond cooling down the system) is also of interest in
soft and biomatter systems (where cooling is not even be possible).
In these situations the second law only yields constraints for the
average behavior, whereas individual realizations may extract work
from the heat bath thereby consuming instead of producing entropy
\cite{ECM}. The characterization of work and heat distributions in
fluctuating non-equilibrium situations has benefited from recent
progress in statistical mechanics centered around the so-called work
and fluctuation theorems, see \cite{BoKu,Jarzynski1997,Jar2} and
\cite{GaCo,Kurchan,JarFT,Crooks,Seifert} respectively.

A trademark of these identities are exponential averages which are
dominated by the large deviation properties of the underlying
probability distributions. When implementing, e.g., the Jarzynski
equation, $\mathrm{e}^{-\beta \Delta F}=\langle \mathrm{e}^{-\beta W}
\rangle$ to estimate the free energy difference $\Delta F$ from the
distribution $P(W)$ of the work, where $\beta=\frac{1}{kT}$ is the
inverse temperature, high accuracy is hindered by the fact that $P(W)$
markedly differs from the distribution
$\tilde{P}(W)\propto\mathrm{e}^{-\beta W} P(W)$ of {\em dominant}
contributions to the exponential average \cite{Jar06}. It is then
relevant to ask for the protocol $\lambda(t)$ that minimizes the
mismatch between $P(W)$ and $\tilde{P}(W)$. A convenient measure for
the similarity between the two distributions is the so-called
Kullback-Leibler divergence \cite{CoTh}
\begin{align}
  D_{\text{KL}}(P||\tilde{P}) =\int \mathrm{d}W P(W) \ln
  \frac{P(W)}{\tilde{P}(W)} =\beta \langle W \rangle - \beta \Delta F.
\end{align}
Consequently one has to search for the protocol that minimizes the
average work. This statement can be sharpened in the linear response
regime, where the work fluctuations are proportional to the dissipated
work \cite{Hermans1991,Jarzynski1997}, $\langle\Delta
W^2\rangle=\frac{2}{\beta}\langle W_{\text{diss}}\rangle$, as proven
in \cite{SpeckSeifert2004}. If one is interested in extracting a sharp
value for the mean work that is required, e.g.\ for folding a protein,
it is desirable to drive the folding according to a protocol that
minimizes the fluctuations of the work, $\langle\Delta
W^2\rangle=\langle W^2\rangle-\langle
W\rangle^2=\frac{2}{\beta}\langle W_{\text{diss}}\rangle$, resulting
again in the protocol that minimizes the average work.  (Far from
equilibrium the connection between work fluctuations and dissipation
is more complicated and no general and precise relation between the
two is known.)

In the present paper we investigate the uniqueness and general
properties of optimal protocols $\lambda(t)$ which minimize the
average work necessary to accomplish a given isothermal transition
between two equilibrium states. We first rederive the results obtained
in \cite{SchmiedlSeifert2007} for linear systems.  We then consider a
non-linear system and study the behavior well beyond the validity of
linear response.

\subsection{Model specifics}
We are seeking for the optimal protocol $\lambda(t)$ that minimizes
the average work in a finite-time process of a small system that is
subject to both, deterministic and stochastic forces.  The former can
be controlled experimentally by the external parameter $\lambda(t)$,
whereas the latter originate from thermal fluctuations of the
environment.  Taking the environment to be a heat bath in thermal
equilibrium at temperature $T$, the stochastic forces are modeled via
a Gaussian white noise, $\sqrt{\frac{2}{\beta}}\zeta(t)$, that is
characterized by its vanishing ensemble average,
$\langle\zeta(t)\rangle=0$, and by the absence of any
time-correlations, $\langle\zeta(t)\zeta(t')\rangle=\delta(t-t')$.
Concerning the deterministic forces, we assume that Stokes friction is
present, $F=-\gamma\dot{x}$, where the force of friction is
proportional to the velocity $\dot{x}$ of the system along its
trajectory $x(t)$. The proportionality constant $\gamma$ is the
friction coefficient. All the other deterministic forces are assumed
to be conservative.

For systems on the nanometer scale-size, e.g.\ biological systems on
the cellular or subcellular level and single molecule experiments, the
stochastic forces exceed the inertial forces by far.  Neglecting the
acceleration term $m\ddot{x}$ in the overdamped motion, where $m$ is
the mass that is accelerated, the microscopic dynamics is described by
the Langevin equation
\begin{align}
  \gamma\dot{x}=-\nabla_x
  V(x(t),\lambda(t))+\sqrt{\tfrac{2}{\beta}}\zeta(t),
  \label{eq:1}
\end{align}
where $V=V(x,\lambda)$ is the potential of the conservative
deterministic forces whose time-dependence is attributed to the
control parameter $\lambda(t)$.  Rescaling the time, we can set the
friction constant to unity, $\gamma=1$.

The time evolution of the probability distribution $p(x,t)$ to observe
the system at position $x$ at time $t$ is governed by the
Fokker-Planck equation
\begin{align}
  \partial_t p(x,t)=\nabla_x(p\nabla_x V)+\tfrac{1}{\beta}\nabla_x^2
  p.
  \label{eq:2}
\end{align}
Starting in thermal equilibrium, the initial canonical distribution is
\begin{align}
  p(x,0)=\frac{\mathrm{e}^{-\beta V(x,\lambda_0)}}{Z_0},
  \label{eq:3}
\end{align}
where the normalization constant $Z_0$ is the partition function.
According to the stochastic forces, any trajectory $[x(t)],\ 0\le t\le
t_f,$ is possible and occurs with probability
\begin{align}
  p[x(\cdot),\lambda(\cdot)] ={\cal
    N}\exp(-\int_0^{t_f}\mathrm{d}t'L(x,\dot{x},\lambda)),
  \label{eq:4}
\end{align}
where the integrand in the exponent of the probability functional
reads
\begin{align}
  L=\tfrac{\beta}{4}(\dot{x}+\nabla_x V)^2-\tfrac{1}{2}\nabla_x^2 V
  \label{eq:5}
\end{align}
and ${\cal N}$ is a normalization constant.  Each realization of the
process requires its specific amount of work
\begin{align}
  W[\lambda(\cdot),x(\cdot)]=V|_{t'=0^-}^{t'=0^+}
  +\int_{0^+}^{t_f^-}\mathrm{d}t'\dot{\lambda} \frac{\partial
    V}{\partial\lambda}+V|_{t'=t_f^-}^{t'=t_f^+},
  \label{eq:6}
\end{align}
where we take possible jumps at the beginning and at the end of the
protocol explicitly into account, e.g.
\begin{align}
  V|_{t'=t_f^-}^{t'=t_f^+}=V(x(t_f),\lambda(t_f^+))-V(x(t_f),\lambda(t_f^-)).
\end{align}
Averaging $W[\lambda(\cdot),x(\cdot)]$ over the initial distribution
and the noisy history the average work
\begin{multline}
  \langle W[\lambda(\cdot)]\rangle =\int_{\mathds{R}}\mathrm{d}x_0
  p(x_0,0)\int_{\mathds{R}}\mathrm{d}x_f \times \\ \times
  \int_{(x_0,0)}^{(x_f,t_f)}\mathscr{D}x(\cdot)
  p[x(\cdot),\lambda(\cdot)]W[\lambda(\cdot),x(\cdot)],
  \label{eq:7}
\end{multline}
becomes a functional of the protocol $[\lambda(t)],\ 0\le t\le t_f,$
according to which the parameter $\lambda(t)$ is varied from its
initial value $\lambda_0$ at $t=0$ to its final value $\lambda_f$ at
$t=t_f$.

The optimal protocol is found by solving the non-local Euler-Lagrange
equation
\begin{align}
  0=\frac{\delta}{\delta\lambda(t)}\langle W[\lambda(\cdot)]\rangle,
  \label{eq:8}
\end{align}
where the variation reads \cite{SchmiedlSeifert2007}
\begin{multline}
  \frac{\delta}{\delta\lambda(t)}\langle W[\lambda(\cdot)]\rangle
  =-\frac{\mathrm{d}}{\mathrm{d}t} \langle\frac{\partial
    V}{\partial\lambda}\rangle_{|t}
  +\dot{\lambda}\langle\frac{\partial^2
    V}{\partial\lambda^2}\rangle_{|t}
  \\
  +\int_t^{t_f^-}\mathrm{d}t' \dot{\lambda}\langle-\frac{\partial
    L}{\partial\lambda}_{|t} \frac{\partial
    V}{\partial\lambda}_{|t'}\rangle +\langle-\frac{\partial
    L}{\partial\lambda}_{|t} V|_{t'=t_f^-}^{t'=t_f^+}\rangle.
  \label{eq:9}
\end{multline}

\subsection{Known results}
Schmiedl and Seifert studied the motion of a colloidal particle in an
optical tweezer \cite{SchmiedlSeifert2007}.  For two cases, namely for
varying the position and the strength of the trap, respectively, they
could express the mean work as a local functional of one variable.
This allowed them to find the optimal protocol that minimizes the mean
work.  As a surprising result Schmiedl and Seifert found the optimal
protocol to jump at the beginning and at the end of the process, i.e.\
at $t=0$ and $t=t_f$, whereas in between the optimal protocol varies
smoothly.  The initial jump can be interpreted as an immediate jump
from equilibrium to a stationary state in order not to loose valuable
time and the final jump allows a slower driving of the system at
earlier times. It is worth noticing that the optimal protocol is
unique in the two cases studied by Schmiedl and Seifert. Both, the
uniqueness of the protocol and the possibility to express the mean
work as a local functional of one variable, result from the linearity
of the systems considered.

Below, we show how these results can be rederived by an explicit
solution of the Euler-Lagrange equation (\ref{eq:8}), (\ref{eq:9}).
For a generic non-linear problem such an analytic solution seems
impossible. We therefore analyze a simple non-linear system
numerically and discuss the new features of the optimal protocol
arising in this case.

\section{The analytic solution}
In the following we demonstrate that an analytic solution of the
Euler-Lagrange equation (\ref{eq:8}), (\ref{eq:9}) is possible if, as
a necessary condition, the underlying Langevin equation (\ref{eq:1})
can be explicitly integrated.  The main trick consists of combining
the Euler-Lagrange equation with its derivatives in such a way that
the integrals cancel out.  Below, this is carried through explicitly
for the two cases studied by Schmiedl and Seifert, i.e.\ for the
stochastic motion of a colloidal particle in an optical tweezer.

\subsection{Case study I}
Dragging a colloidal particle through a viscous fluid by an optical
tweezer with harmonic potential
\begin{align}
  V(x,\lambda)=\tfrac{1}{2}(x-\lambda)^2,
  \label{eq:10}
\end{align}
where the focus of the optical tweezer is moved according to a
protocol $\lambda(t)$ from $\lambda(0^-)=\lambda_0$ to
$\lambda(t_f^+)=\lambda_f$ in a finite time $t_f$, the expressions
appearing in the variation of the average work (\ref{eq:9}) can be
computed,
\begin{align}
  &-\frac{\partial
    L}{\partial\lambda}_{|t}=\sqrt{\tfrac{\beta}{2}}\zeta(t),
  \\
  &x(t)=\mathrm{e}^{-t}(x(0)+\int_{0}^{t}\mathrm{d}\tau
  (\lambda(\tau)+\sqrt{\tfrac{2}{\beta}}\zeta(\tau))\mathrm{e}^{\tau}),
  \\
  &\langle\zeta(t)x(t')\rangle
  =\sqrt{\tfrac{2}{\beta}}\mathrm{e}^{-(t'-t)}\Theta(t'-t),
\end{align}
and the Euler-Lagrange equation becomes
\begin{multline}
  0=-\mathrm{e}^{-t}\lambda(0^-)
  -\mathrm{e}^{-t}\int_{0}^{t}\mathrm{d}t'\lambda(t')\mathrm{e}^{t'}
  \\
  +2\lambda(t)
  -\mathrm{e}^{t}\int_{t}^{t_f}\mathrm{d}t'\lambda(t')\mathrm{e}^{-t'}
  -\lambda(t_f^+)\mathrm{e}^{-(t_f-t)}
  \label{eq:11}
\end{multline}
for all $t\in(0,t_f)$.

Differentiating the Euler-Lagrange equation twice it is almost
identically reproduced.  From the difference between the
Euler-Lagrange equation and its second derivative follows a simple
differential equation for the optimal protocol,
\begin{align}
  \ddot{\lambda}(t)=0 \quad \forall t\in(0,t_f).
  \label{eq:12}
\end{align}
Inserting the solution of (\ref{eq:12}), $\lambda(t)=at+b$, back into
(\ref{eq:11}) and setting $\lambda_0=0$ yields the integration
constants to be equal to
\begin{align}
  a=b=\frac{\lambda_f}{t_f+2},
  \label{eq:13}
\end{align}
resulting in the optimal protocol to be
\begin{align}
  \lambda(t) =
  \begin{cases}
    0 & , \quad t \le 0 \, , \\
    \tfrac{ \lambda_f}{ t_f + 2}( t + 1) & , \quad 0 < t < t_f \, , \\
    \lambda_f & , \quad t \ge t_f \, ,
  \end{cases}
  \label{eq:14}
\end{align}
in agreement with the result of Schmiedl and Seifert
\cite{SchmiedlSeifert2007}.

\subsection{Case study II}
Varying the strength of the trap,
\begin{align}
  V(x,\lambda)=\tfrac{1}{2}\lambda x^2,
  \label{eq:15}
\end{align}
with $\lambda(0^-)=\lambda_0>0$ and $\lambda(t_f^+)=\lambda_f>0$ as
boundary conditions, the expressions appearing in the variation of the
average work (\ref{eq:9}) can be computed again,
\begin{align}
  &-\frac{\partial
    L}{\partial\lambda}_{|t}=-\sqrt{\tfrac{\beta}{2}}\zeta(t)x(t)
  +\frac{1}{2},
  \\
  &x(t)=\mathrm{e}^{-\Lambda(t)}(x(0)+\sqrt{\tfrac{2}{\beta}}
  \int_{0}^{t}\mathrm{d}\tau\mathrm{e}^{\Lambda(\tau)}\zeta(\tau)),
  \\
  &\langle x^2(t)\rangle=\mathrm{e}^{-2\Lambda(t)} (\langle
  x^2(0)\rangle+\tfrac{2}{\beta}
  \int_{0}^{t}\mathrm{d}\tau\mathrm{e}^{2\Lambda(\tau)}),
  \\
  &\tfrac{\mathrm{d}}{\mathrm{d}t}\langle\tfrac{1}{2}x^2(t)
  \rangle=-\lambda(t)\langle x^2(t)\rangle+\tfrac{1}{\beta},
\end{align}
with
\begin{align}
  \Lambda(t):=\int_{0}^{t}\mathrm{d}t'\lambda(t')
\end{align}
and
\begin{align}
  \langle x^2(0)\rangle=\tfrac{1}{\beta\lambda_0}.
\end{align}
Using Stratonovich calculus, we have
\begin{multline}
  \langle\zeta(t)x(t)x^2(t')\rangle
  \\
  =\sqrt{\tfrac{2}{\beta}}\tfrac{1}{\beta}
  \mathrm{e}^{-2\Lambda(t')}(\tfrac{5}{2\lambda_0}
  +\int_{0}^{t'}\mathrm{d}\tau\mathrm{e}^{2\Lambda(\tau)}
  +4\int_{0}^{t}\mathrm{d}\tau\mathrm{e}^{2\Lambda(\tau)})
\end{multline}
for $t'>t>0$, and the Euler-Lagrange equation reads
\begin{align}
  0=\frac{\delta}{\delta\lambda(t)}\langle
  W\rangle=\frac{1}{\beta}(AB-1) =:\frac{1}{\beta}C,
  \label{eq:16}
\end{align}
where $A$ and $B$ are abbreviations for
\begin{multline}
  A:=\lambda(t)\mathrm{e}^{-2\Lambda(t)}
  -\int_{t}^{t_f^-}\mathrm{d}t'\dot{\lambda}(t')
  \mathrm{e}^{-2\Lambda(t')}
  \\
  -(\lambda_f-\lambda(t_f^-))\mathrm{e}^{-2\Lambda(t_f)}
  \label{eq:17}
\end{multline}
and
\begin{align}
  B:=\frac{1}{\lambda_0}+2\int_{0}^{t}\mathrm{d}t''\mathrm{e}^{2\Lambda(t'')}.
  \label{eq:18}
\end{align}

Note that $C = A B - 1, \ \dot{C} = \dot{A} B + A \dot{B}, \ \ddot{C}
= \ddot{A} B + 2 \dot{A} \dot{B} + A \ddot{B}, \ \ldots \ $ are
complicated, because with $A$ and $B$ they contain integrals and
exponentials of integrals.
 
With $\dot{A} = ( 2 \dot{ \lambda} - 2 \lambda^2) \mathrm{e}^{ -2
  \Lambda}$ and $\dot{B} = 2 \mathrm{e}^{ 2 \Lambda}$, the functions
$\dot{A}\dot{B},\ \ddot{A}\dot{B}, \ \dot{A}\ddot{B},\
\ddot{A}\ddot{B},\ \ldots \ $ do not contain any integral.  Our goal
is to express $C$ by these simpler functions.
 
From $\ddot{A} \dot{C} - \dot{A} \ddot{C} = A ( \ddot{A} \dot{B} -
\dot{A} \ddot{B}) - 2 \dot{A}^2 \dot{B}$ and $\ddot{B} \dot{C} -
\dot{B} \ddot{C} = B ( \ddot{B} \dot{A} - \dot{B} \ddot{A}) - 2
\dot{B}^2 \dot{A}$ follows an ordinary differential equation for $C$,
\begin{multline}
  C = A B - 1
  \\
  = \frac{ ( \ddot{A} \dot{C} - \dot{A} \ddot{C} + 2 \dot{A}^2
    \dot{B}) ( \ddot{B} \dot{C} - \dot{B} \ddot{C} + 2 \dot{B}^2
    \dot{A})} { ( \ddot{A} \dot{B} - \dot{A} \ddot{B}) ( \ddot{B}
    \dot{A} - \dot{B} \ddot{A}) } - 1 .
  \label{eq:20}
\end{multline}

Remember that we are interested in extremizing the average work
\begin{align}
  \frac{ 1}{ \beta} C \equiv \frac{ \delta}{ \delta \lambda(t)}
  \langle W \rangle = 0 \quad \forall \, t \in ( 0, t_f) .
  \label{eq:21}
\end{align}
Consequently, $C$ and its derivatives have to vanish identically.
This yields
\begin{align}
  0 = \frac{ 4 \dot{A}^3 \dot{B}^3}{ ( \ddot{A} \dot{B} - \dot{A}
    \ddot{B}) ( \ddot{B} \dot{A} - \dot{B} \ddot{A})} - 1
  \label{eq:22}
\end{align}
or
\begin{multline}
  0 = 4 ( \dot{A} \dot{B})^3 + ( \ddot{A} \dot{B} - \dot{A}
  \ddot{B})^2
  \\
  = 16 ( \ddot{ \lambda}^2 - 12 \lambda \dot{ \lambda} \ddot{ \lambda}
  + 8 \lambda^3 \ddot{ \lambda} + 16 \dot{ \lambda}^3 - 12 \lambda^2
  \dot{ \lambda}^2) .
  \label{eq:23}
\end{multline}
The latter is an ordinary differential equation for the optimal
protocol.  Using Lie symmetries \cite{Hydon2000} it can be decomposed
and integrated resulting in $\lambda=\lambda_{1}$ and
$\lambda=\lambda_{2,3}$ with
\begin{align}
  \lambda_{1}(t) = \frac{ 1}{ b - t} \quad \text{ and } \quad
  \lambda_{2,3}(t) = \frac{ a - c( 1 + ct)}{ ( 1 + ct)^2} .
  \label{eq:24}
\end{align}
If plugged into the Euler-Lagrange equation (\ref{eq:16}),
$\lambda_{1}(t)$ fails to describe the solution for positive
$\lambda_0$.  $\lambda_{2,3}(t)$ solve the Euler-Lagrange equation
provided the integration constants are $a=\lambda_0$ and
\begin{align}
  c = \frac{ -1 - \lambda_f t_f + \sqrt{ 1 + 2 \lambda_0 t_f +
      \lambda_0 \lambda_f t_f^2}}{ t_f ( 2 + \lambda_f t_f)} .
  \label{eq:25}
\end{align}
The optimal protocol is
\begin{align}
  \lambda(t) =
  \begin{cases}
    \lambda_0 & , \quad t \le 0 \, , \\
    \displaystyle \frac{ \lambda_0 - c( 1 + ct)}{ ( 1 + ct)^2} & ,
    \quad 0 < t < t_f \, , \\
    \lambda_f & , \quad t \ge t_f \, ,
  \end{cases}
  \label{eq:26}
\end{align}
in agreement with the result of Schmiedl and Seifert
\cite{SchmiedlSeifert2007}.

\section{Numerical methods}
We have shown how the Euler-Lagrange equation can be solved
analytically.  Thereby, it was essential that the Langevin equation
can be integrated.  Often, this equation cannot be integrated
analytically.  In the latter case, it is not even possible to express
the Euler-Lagrange equation as an integro-differential equation in
$\lambda$, because the correlation functions cannot be evaluated
explicitly, and one has to resort to numerical methods.

Numerically, we have implemented a Monte Carlo algorithm that
minimizes the average work directly.  For a given protocol the average
work (\ref{eq:7}) is approximated using a finite ensemble of
trajectories that are discretized in time, $\{x_n(t_m)\}$, $n=1\ldots
N$ and $m=0\ldots M$ with $t_0=0$ and $t_M=t_f$.  The initial
distribution $\{x_n(0)\}$ according to (\ref{eq:3}) and the noisy
history $\{\zeta_n(t_m)\}$ for each trajectory of the ensemble are
diced using the Ziggurat method \cite{MarsagliaTsang2000}.  For each
trajectory the Langevin equation (\ref{eq:1}) is integrated according
to the Heun-scheme \cite{Blum1972,GarciaPalaciosLazaro1998}.  Finally,
the optimal protocol is found with the threshold acceptance algorithm
\cite{DueckScheuer1990}.

In the threshold algorithm we approximate the protocol by a polygon
line that connects the $2Q+2$ points
$\{(0,\lambda_{0^-}),(0,\lambda_{0^+}),
(\frac{1}{Q}t_f,\lambda_{1^-}),(\frac{1}{Q}t_f,\lambda_{1^+}),\ldots,$
$(\frac{Q-1}{Q}t_f,\lambda_{(Q-1)^-}),(\frac{Q-1}{Q}t_f,
\lambda_{(Q-1)^+}),(t_f,\lambda_{f^-}),(t_f,\lambda_{f^+})\}$, where
the boundary values are $\lambda_{0^-}=\lambda_0$ and
$\lambda_{f^+}=\lambda_f$, and $Q$ is a positive integer.  The
threshold algorithm is an iterative algorithm that starts with a
random choice of initial values for
$\{\lambda_{0^+},\ldots,\lambda_{f^-}\}$ and computes the
corresponding average work. In each iteration the values
$\{\lambda_{0^+},\ldots,\lambda_{f^-}\}$ are randomly perturbed and
the corresponding average work is compared with the average work of
the best protocol that was found in the previous iterations.  If the
protocol results in an average work that is not deteriorated by more
than some given threshold value, it is used as the best protocol in
the next iteration. Whenever no better protocol is found in some
finite number of iterations, the algorithm lowers the threshold and
continues iterating. If finally the algorithm does not find a better
protocol in some finite number of iterations at zero threshold, it
eliminates intermediate jumps in the protocol provided the protocol
keeps optimal.

It is worth to note that we dice the initial distribution and the
noisy history only once and reuse these values in any iteration of the
algorithm.  This is not just to speed up the numerics, it is necessary
for the algorithm to converge.

Using up to about $N=7000$ trajectories discretized into $M=154$
time-steps and approximating the protocol using a few points for the
polygon, $2Q+2=44+2$, the algorithm is reasonably fast and can easily
be run on a single processor.  Starting from a random initial
protocol, it typically converges in less than $30000$ iterations.
Depending on the parameter values, it finds the optimal protocol
within a few seconds or minutes, but for some parameter values it runs
for several hours. The result is a first approximation of the optimal
protocol.

In order to increase the numerical resolution, we rerun the algorithm
using up to $N=50000$ trajectories discretized into $M=700$
time-steps. Starting from the previously found first approximation of
the optimal protocol, but now approximating the protocol using
$2Q+2=200+2$ polygon points, the algorithm typically converges in less
than $70000$ iterations. The required CPU-time of the rerun is by a
factor of $\sim40$ larger, because of the increased number of
trajectories and time-steps involved.

With the numerics at hand, we study a further example.

\subsection{Case study III}
Consider the stochastic motion of a small dipole in a viscous liquid
driven by an external field where the direction of the external field
is changed according to a protocol $\lambda(t)$. The protocol may
start in any arbitrary initial direction $\lambda_0$ at time $t=0$ and
ends in the final direction $\lambda_f=\lambda_0+\Delta\lambda$ at
time $t=t_f$. For notational simplicity, we regard one degree of
freedom and explore the stochastic motion
\begin{align}
  \dot{x}=-\frac{\partial}{\partial x}
  V(x(t),\lambda(t))+\sqrt{2D}\zeta(t)
\end{align}
in the potential
\begin{align}
  V(x,\lambda)=-H\cos(x-\lambda),
  \label{eq:27}
\end{align}
where the angle $x$ characterizes the orientation of the dipole.  In
the numerics, we set the amplitude of the field to unity, $H=1$, and
stop the protocol at $t_f=1$.

Being subject to two time-scales, the relaxation time and the time for
driving the system, the motion of the dipole depends qualitatively on
the diffusion constant, $D=\frac{1}{\beta}$, which we take as an
additional constant parameter into account.

\begin{figure}
  \includegraphics{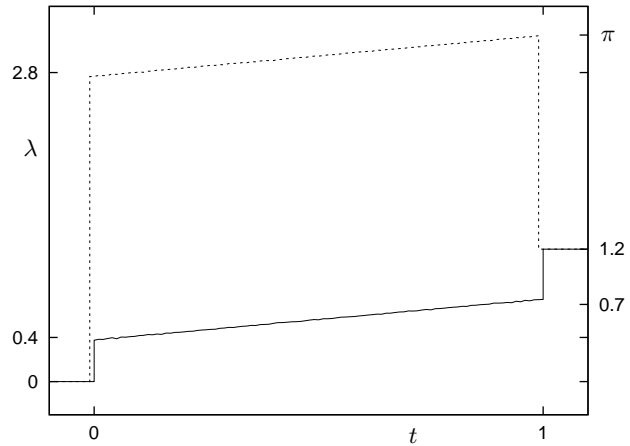}
  \caption{\label{fig:1}The optimal protocol (solid line) that
    minimizes the average work for $V=-\cos(x-\lambda)$ with
    $D=10^{-5}$ and $\Delta\lambda=1.2$ is unique and jumps twice. But
    it is interesting to see that there exists a suboptimal protocol
    (dotted line) whose average work, $\langle W\rangle=0.4620$, is
    close to the average work of the optimal protocol, $\langle
    W\rangle=0.4619$. For better visualization we have shifted the
    suboptimal protocol slightly in time.}
\end{figure}

\begin{figure}
  \includegraphics{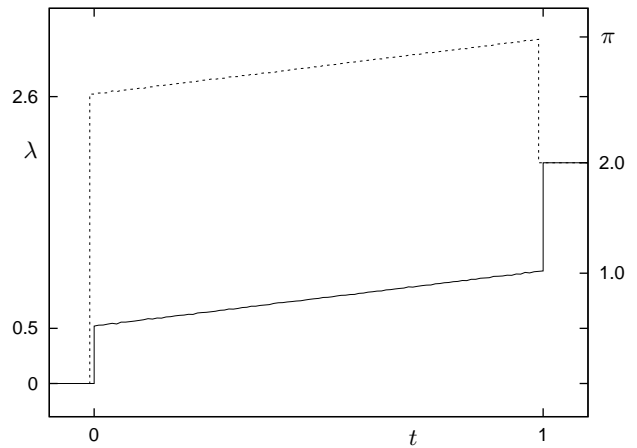}
  \caption{\label{fig:2}The optimal protocol (solid line) that
    minimizes the average work for $V=-\cos(x-\lambda)$ with
    $D=10^{-5}$ and $\Delta\lambda=2.0$ requires an average work of
    $\langle W\rangle=1.1791$.  With $\langle W\rangle=1.1792$, the
    average work of the suboptimal protocol (dotted line) almost
    approaches that of the optimal protocol.}
\end{figure}

\begin{figure}
  \includegraphics{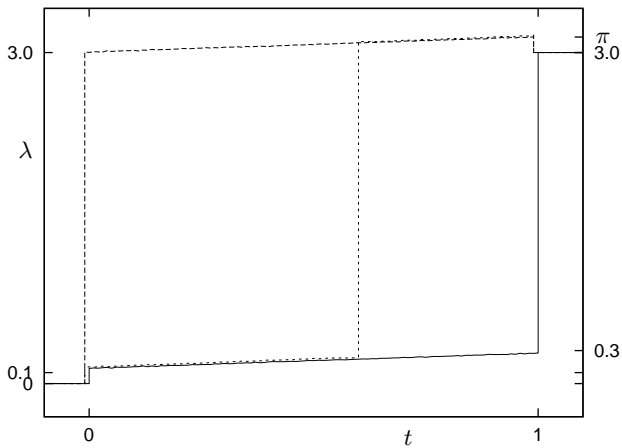}
  \caption{\label{fig:3}For $V=-\cos(x-\lambda)$ with $D=10^{-5}$ and
    $\Delta\lambda=3.0$ there is a whole family of optimal protocols.
    Two of them jump twice (solid and dashed lines).  All the other
    optimal protocols jump at least three times.  One of them is
    displayed by the dotted line.  The average work for any member of
    the family of optimal protocols is $\langle W\rangle=1.9802$.}
\end{figure}

\begin{figure}
  \includegraphics{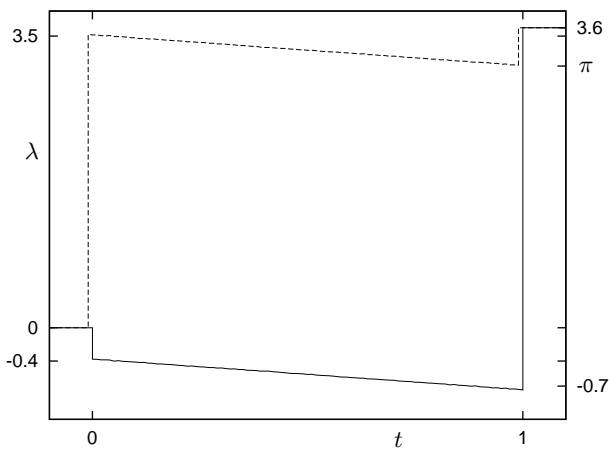}
  \caption{\label{fig:4}Two members of the family of optimal protocols
    are displayed for $V=-\cos(x-\lambda)$ with $D=10^{-5}$ and
    $\Delta\lambda=3.6$. The two displayed protocols jump twice. All
    the other optimal protocols jump at least three times. The average
    work is $\langle W\rangle=1.8130$.}
\end{figure}

\begin{figure}
  \includegraphics{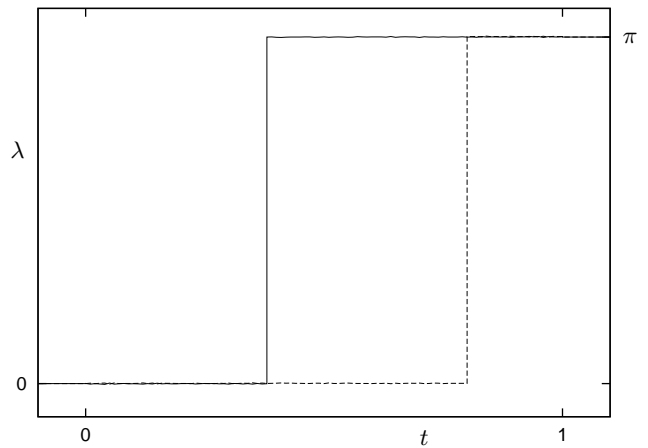}
  \caption{\label{fig:5}For $V=-\cos(x-\lambda)$ with
    $\Delta\lambda=\pi$ and $D=10^{-5}$ the initial and the final jump
    vanish and the optimal protocol can degenerate to one single jump
    that may happen at any arbitrary time, $0\le t\le t_f$. Two
    members of the family of optimal protocols are displayed. The
    average work is $\langle W\rangle=1.99999$.}
\end{figure}

\begin{figure}
  \includegraphics{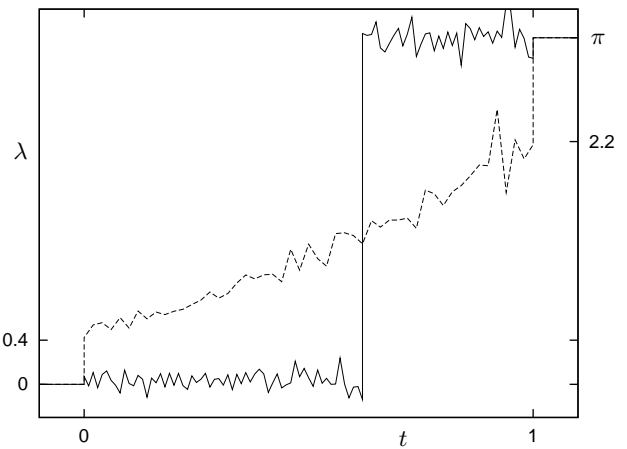}
  \caption{\label{fig:6}Optimal protocols that minimize the average
    work for $V=-\cos(x-\lambda)$ with $\Delta\lambda=\pi$.  For
    $D=0.5$ there is a family of optimal protocols.  The solid line
    mimics one member of this family. In contrast, for $D=0.7$ the
    optimal protocol is unique and jumps twice as mimiced by the
    dashed line.}
\end{figure}

Simulating the stochastic motion of the dipole, we find optimal
protocols as displayed in figures \ref{fig:1} to \ref{fig:6}.

Let us first discuss the numerical solutions with the diffusion
constant kept small, $D=10^{-5}$.  If
$\Delta\lambda=\lambda_f-\lambda_0$ is also small, the potential can
be Taylor expanded around its minimum and is approximately equal to
that of the moving laser trap of case study I. The solid line in
figure \ref{fig:1} displays the optimal protocol for
$\Delta\lambda=1.2$.

If the angle $\Delta\lambda$ increases, the dynamics starts to
experience the non-linearities of the potential. For e.g.\
$\Delta\lambda=2.0$ the final jump becomes much larger than the
initial jump, see the solid line in figure \ref{fig:2}.

In rare cases, the numerical algorithm converges to another protocol,
see the dotted lines in figures \ref{fig:1} and \ref{fig:2}.  The
values of the average work tell us that these latter protocols are
slightly suboptimal, $\langle W\rangle=0.4620$ and $\langle
W\rangle=1.1792$, while the optimal protocols require $\langle
W\rangle=0.4619$ and $\langle W\rangle=1.1791$, respectively.
However, we cannot really decide whether the dotted line in figure
\ref{fig:2} is a suboptimal protocol or whether it is optimal, because
the differences in the average work are close to the resolution of our
numerical procedure.

If the angle $\Delta\lambda$ is further increased, to e.g.\
$\Delta\lambda=3.0$, the previously suboptimal protocol becomes
optimal. Both protocols, indicated by the solid and the dashed lines
in figure \ref{fig:3} yield the same value for the average work,
$\langle W\rangle=1.9802$.  The two optimal protocols differ by the
sizes of their initial and final jumps, whereas their slopes are
identical for $0<t<t_f$.

Being the result of a numerical procedure, we can never claim in a
strict mathematical sense that two protocols are both optimal.  In
practice, however, it is impossible to distinguish the case of two
optimal protocols from that of two protocols with extremely close work
values.

\begin{figure}
  \includegraphics{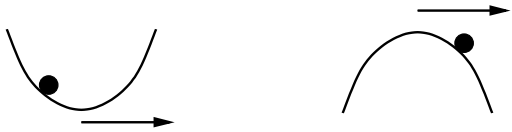}
  \caption{\label{fig:7}The two different states that drive the
    orientation of the dipole. The left figure displays the stable
    state where the system is pulled by the minimum of the potential.
    The right figure displays the unstable state where the system is
    pushed by the maximum of the potential.}
\end{figure}

The existence of two optimal protocols results from the symmetry in
the potential, $-\cos(x-\lambda)=\cos(\pi-(x-\lambda))$.  The optimal
protocol, as indicated by the solid line in figure \ref{fig:3}, jumps
at $t=0$ to a stable state that pulls the dipole towards a new
orientation, see figure \ref{fig:7} (left).  This protocol needs most
of its average work in the final jump.  The other optimal protocol,
displayed by the dashed line in figure \ref{fig:3}, requires most of
the average work in the initial jump.  Immediately after this jump,
the system is in a state where the dipole is pushed by the potential,
see figure \ref{fig:7} (right).  Because of the symmetry in the
potential between its minimum and its maximum, the slopes of the two
optimal protocols are the same for $0<t<t_f$.

Beyond these two optimal protocols, for $D$ small and $\Delta\lambda$
close to $\pi$, there is a whole family of optimal protocols.  Any
protocol of this family starts with an initial jump that brings the
system to its pulled or pushed state, respectively.  At any arbitrary
time the protocol can jump again and the system switches between the
pulled and the pushed state.  Such a protocol is displayed by the
dotted line in figure \ref{fig:3}.  Finally, the optimal protocol can
jump repeatedly between the the pulled and the pushed state allowing
for any number of jumps.

For $\Delta\lambda$ approaching $\pi$, the size of the initial and/or
the final jump, and the slope of the optimal protocols decrease
towards zero, cf.\ figures \ref{fig:2} and \ref{fig:3}.

Increasing $\Delta\lambda$ further such that it exceeds $\pi$, the
initial and/or the final jump, and the slope of the optimal protocols
change sign, cf.\ figures \ref{fig:3} and \ref{fig:4}.

Due to the change in signs and because of the symmetry and periodicity
of the potential, we know that for $\Delta\lambda=\pi$ the initial
and/or the final jump, and the slope of the optimal protocols vanish
identically, and we conclude the following:

For $D$ small and $\Delta\lambda$ close to $\pi$, the optimal protocol
jumps at least two times, except if $\Delta\lambda=\pi$ holds
identically. In the latter case, the initial and/or the final jump
vanish and the whole protocol can degenerate to one single jump that
may happen at any arbitrary time, $0\le t\le t_f$.  Two members of the
family of optimal protocols are displayed in figure \ref{fig:5}.

It is interesting to consider also the variances of the work for the
different optimal protocols. While they give the same value for the
average work, the variances of the work differ slightly among the
family members. The optimal protocol that pulls the system all the
time in the stable state, see figure \ref{fig:7} (left), yields the
smallest variance for the work, whereas the optimal protocol that
pushes the system all the time in the unstable state, figure
\ref{fig:7} (right), yields the largest variance for the work. Note
that different values for the variances of work for protocols that
yield the same average work imply that we are not in the linear
response regime.

Because of the limited numerical resolution, the slight differences in
the variances of the work have to be treated with some care.
Nevertheless, the fact that we are well beyond the linear response
regime can clearly be demonstrated by checking whether a central
identity of linear response theory, $\langle\Delta
W^2\rangle=2D\langle W_{\text{diss}}\rangle$
\cite{Hermans1991,Jarzynski1997,SpeckSeifert2004}, is violated.  For
example, the optimal protocol for $\Delta\lambda=2.0$, see figure
\ref{fig:2}, results in $2D\langle
W_{\text{diss}}\rangle/\langle\Delta W^2\rangle=1.76\not=1$, and the
optimal protocols for $\Delta\lambda=\pi$, see figure \ref{fig:5},
result in $2D\langle W_{\text{diss}}\rangle/\langle\Delta W^2\rangle
\approx10^5\not=1$.

Yet, we have kept the diffusion constant $D$ small. Increasing the
diffusion constant increases the noise, see the scrambled lines that
mimic the optimal protocols in figure \ref{fig:6}. In principle, we
can get rid of the fluctuations in taking the average over larger
ensembles, but we have decided to use only $50000$ trajectories for
the ensembles in order to keep the CPU-time reasonable.  Moreover, in
computing the optimal protocol for $D=0.7$ we have reduced the
resolution to $2Q+2=100+2$ points, otherwise the CPU-time would exceed
two weeks.

Beyond the quantitative change in the amplitude of the fluctuations,
there is also a qualitative change in the behavior of the system. As
discussed above, if the direction of the external field is changed by
the angle $\Delta\lambda=\pi$ in time $t_f=1$ and the diffusion
constant is small, there exist optimal protocols that consist of only
one single jump that may happen at any arbitrary time. The solid line
in figure \ref{fig:6} shows that such protocols are optimal up to
$D=0.5$.

If the diffusion becomes larger, the previously optimal protocols that
were allowed to jump several times, get suboptimal.  Instead, a
different protocol becomes optimal. The latter is unique.  The
transition happens somewhere below $D=0.7$. The dashed line in figure
\ref{fig:6} mimics the optimal protocol for $D=0.7$, having two jumps,
one at $t=0$ and one at $t=t_f=1$.

\begin{figure}
  \includegraphics{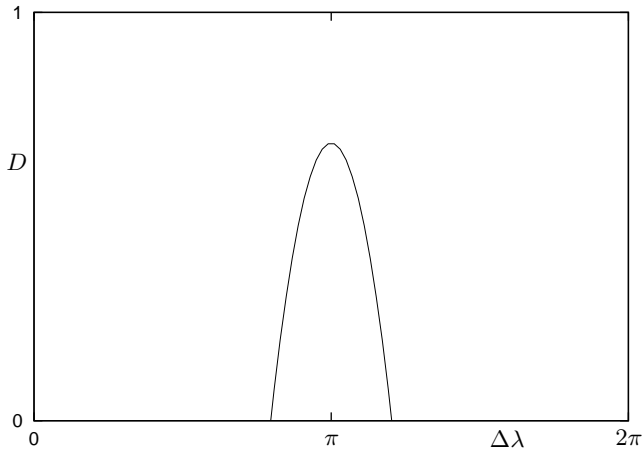}
  \caption{\label{fig:8}There are two regions in the
    $\Delta\lambda$-$D$-plane.  For $\Delta\lambda$ near $\pi$ and $D$
    small, the optimal protocol occurs in a family and may jump
    several times.  For $\Delta\lambda$ far away from $\pi$ or for $D$
    large, the optimal protocol is unique and jumps twice.}
\end{figure}

There are hence two regions in the $\Delta\lambda$-$D$-plane. These
are shown qualitatively in figure \ref{fig:8}. One region is located
around $\Delta\lambda\approx\pi$ and $D$ small. In this region, the
optimal protocol occurs in a family and may jump several times. In the
other region, i.e.\ for $\Delta\lambda$ far away from $\pi$ or for $D$
large, the optimal protocol is unique and jumps twice.

If one comes close to the curve that separates the two regions in the
$\Delta\lambda$-$D$-plane, it becomes hard to decide numerically which
protocols are optimal and which are suboptimal, because the values of
their average work approach each other. Being blurred by the noisy
history of the trajectories, it becomes quite impossible to determine
the exact curve that separates the two regions in the
$\Delta\lambda$-$D$-plane. For this reason, we show in figure
\ref{fig:8} only a crude approximation of this curve.

\section{Conclusion}
The central subject of this article was to present a guide according
to which the non-local Euler-Lagrange equation, a non-linear
integro-differential equation of correlation functions, can be solved
in order to find the optimal protocol of an external control parameter
that minimizes the average work for driving a small system from one
given equilibrium to another in finite time.

Studying the stochastic motion of a dipole where the direction of the
external field is varied according to some protocol, we have found as
a surprise that the optimal protocol may not be unique. For suitable
parameter values the protocol can jump several times.

The reason for the non-uniqueness of the optimal protocol results from
a symmetry in the potential that allows the dominant trajectories to
be near the unstable maximum of the potential, figure \ref{fig:7}
(right), as an alternative to the stable solution, where the
trajectories are near the minimum of the potential, figure \ref{fig:7}
(left).

If the diffusion becomes large, the optimal protocol becomes unique.

It is tempting to speculate whether biological systems on the cellular
and subcellular level benefit from non-unique optimal protocols.  Just
imagine a protocol to be a triggering mechanism for a cellular process
to be activated. It can be optimal without the need of exact timing,
allowing the biological system to react spontaneously on the
environment.

\section*{Acknowledgments}
We thank Tim Schmiedl and Udo Seifert for useful correspondence.

\end{document}